\documentstyle[aps,floats]{revtex}
\begin{document}  
\twocolumn[\hsize\textwidth\columnwidth\hsize\csname 
           @twocolumnfalse\endcsname
\title{Gravitational waves from inspiraling compact binaries: \\
       Accuracy of the post-Newtonian waveforms}  
\author{Eric Poisson} 
\address{Department of Physics, University of Guelph, Guelph, Ontario,
         N1G 2W1, Canada} 
\date{January 12, 1998}
\maketitle

\begin{abstract} 
\widetext
The accuracy of the post-Newtonian waveforms, both in ``standard'' and
``Pad\'e'' form, is determined by computing their matched-filtering
overlap integral with a reference waveform obtained from black-hole
perturbation theory.
\end{abstract} 
\vskip 2pc]

\narrowtext

\section{Introduction}

The need for very accurate model waveforms to search and measure 
gravitational-wave signals from inspiraling compact binaries was
recognized less than six years ago \cite{CutlerEtal}. During this
time, post-Newtonian (PN) calculations of binary waveforms have 
been pushed to high order \cite{BIWW,Blanchet}. Currently, the 
waveform is known to 2.5PN order beyond the quadrupole-formula 
result, and the 3PN waveform appears to be within reach. 
An ever important question is: at which order in the post-Newtonian
expansion can the waveform be considered to be sufficiently
accurate? In the absence of an exact representation of the signal 
(which might  eventually be provided by numerical relativity), a
definitive answer to this question remains elusive.

An alternative approach to waveform calculations, based of black-hole
perturbation theory, was also pushed forward in the last several 
years \cite{Poisson93,Tanaka}. While post-Newtonian theory is based on an 
assumption of slow orbital motion  but allows for arbitrary mass ratios, 
perturbation theory is based on an assumption of small mass ratios but 
allows for arbitrary velocities. The physical situation is that of a small 
mass orbiting a massive black hole, and the emission of gravitational waves 
is calculated 
by integrating a linear wave equation --- the Teukolsky equation --- in the
black-hole spacetime. In the limit of a vanishing mass ratio, this approach
gives an exact representation of the gravitational-wave signal. When
combined with a slow-motion approximation, perturbation theory returns
an analytic expression for the waves, in terms of a series in powers of 
the orbital velicity. Currently, the perturbation-theory waveform is known
exactly in numerical form, and at 5.5PN order in analytic form \cite{Tanaka}. 

Although the perturbation-theory waveform is of restricted validity, it is 
still our best guide in trying to determine the accuracy of the post-Newtonian 
expansion. In this contribution to these proceedings, this issue will be examined 
in two different contexts. First, supposing that gravitational-wave signals from 
small-mass-ratio binaries can be picked up by such interferometric detectors as 
LIGO/VIRGO, I calculate (Sec.~3) the loss in signal-to-noise ratio incurred by 
using second post-Newtonian waveforms as search templates. My results suggest
that the 2PN waveforms will be acceptably effective for searches. Second, I
review (Sec.~4) the very interesting proposal by Damour, Iyer, and 
Sathyaprakash \cite{DIS} to use Pad\'e approximants to improve 
the convergence of the post-Newtonian series. My results suggest that a 
Pad\'e version of the 3PN waveforms (yet to be produced) will make very 
accurate templates for searches, and possibly also for measurements. 
In Sec.~2, I present the framework used throughout this contribution. 

The work presented here was partially carried out with Serge Droz. Additional
details can be found in \cite{DrozPoisson,Poisson98}.

\section{Preliminaries}

The output of an idealized gravitational-wave detector is written as $o(t) = 
n(t) + s(t)$, where $n(t)$ is stationary Gaussian noise and $s(t)$ is the signal. 
The detector output is analyzed by matched filtering against a bank of templates 
$h(t;\bbox{\lambda},\bbox{\theta})$. Here, $\bbox{\lambda}$ collectively denotes the 
template's {\it kinematical} parameters (arrival time and initial phase), while
$\bbox{\theta}$ collectively represents the {\it dynamical} parameters (the two masses,
assuming that the binary companions are nonrotating). The statistical properties
of the detector noise are summarized by its spectral density, denoted $S_n(f)$,
where $f$ is the frequency. Throughout we use the useful 
approximation \cite{CutlerFlanagan}
$S_n(f) = S_0 \Theta(f-f_{\rm min}) [ (f_0/f)^4 + 2 + 2(f/f_0)^2 ]$, where
the parameters $S_0$, $f_{\rm min}$, and $f_0$ are set by the 
detector. For the initial LIGO, $f_{\rm min} = 40\ \mbox{Hz}$
and $f_0 = 200\ \mbox{Hz}$, while $f_{\rm min} = 10\ \mbox{Hz}$ and 
$f_0 = 70\ \mbox{Hz}$ for the advanced LIGO. The value of $S_0$ is irrelevant
for our purposes.

We define the {\it ambiguity function} ${\cal A}(\bbox{\lambda},\bbox{\theta})$ by
\begin{equation}
{\cal A}(\bbox{\lambda},\bbox{\theta}) = \frac{ (s|h) }{ \sqrt{(s|s) (h|h)} },
\label{1}
\end{equation}
where, for any functions $a(t)$, $b(t)$,
\begin{equation}
(a|b) = 2 \int_0^\infty \frac{\tilde{a}^*(f) \tilde{b}(f) + 
\tilde{a}(f) \tilde{b}^*(f)}{S_n(f)}\, df
\label{2}
\end{equation}
is an inner product in the Hilbert space of gravitational-wave 
signals \cite{CutlerFlanagan}. Here, 
$\tilde{a}(f)$ is the Fourier transform of $a(t)$, and an asterisk denotes complex 
conjugation. The ambiguity function measures the Hilbert-space ``angle'' between
the signal and the template. This ``angle'' varies with the template 
parameters, and we are interested in maximizing the ambibuity function 
over these parameters. We construct the  {\it semi-maximized ambiguity 
function} (SMAF) by maximizing only over the kinematical parameters:
\begin{equation}
\mbox{SMAF}(\bbox{\theta}) = \max_{\bbox{\lambda}} 
{\cal A}(\bbox{\lambda},\bbox{\theta}).
\label{3}
\end{equation}
The SMAF will be computed in Sec.~4. The fully maximized ambiguity
function is Apostolatos' {\it fitting factor} \cite{Apostolatos},
\begin{equation}
\mbox{FF} = \max_{\bbox{\theta},\bbox{\lambda}}
{\cal A}(\bbox{\lambda},\bbox{\theta}).
\label{4}
\end{equation}
The fitting factor is equal to ratio of the {\it actual} signal-to-noise ratio, 
obtained with an imperfect set of templates, to the signal-to-noise ratio that 
{\it would} be obtained if a perfect set of templates were available. Consequently,
the loss in event rate due to template imperfection is $1-\mbox{FF}^3$.
The fitting factor will be computed in Sec.~3.

In the calculations of Secs.~3 and 4, the reference signal will be provided by 
black-hole perturbation theory, as was discussed in Sec.~1. Keeping only the 
dominant mode at twice the orbital frequency, the frequency-domain signal can 
be expressed as \cite{DrozPoisson}
\begin{equation}
\tilde{s}(f) \propto A(v)\, e^{i \Psi(v)},
\label{5}
\end{equation}
where $v = (\pi M f)^{1/3}$ is the orbital velocity ($M$ is the total mass
of the binary system, and $\mu$ will denote the reduced mass); $A(v)$ 
is the signal's amplitude, and $\Psi(v)$ its phase. When using the 
stationary-phase approximation to calculate the Fourier transform,
the phase is found to be given by  
\begin{equation}
\Psi(v) = \frac{5M}{16\mu} \int^v 
\frac{ (v^3-v^{\prime 3}) Q(v') }{ v^{\prime 9} P(v')}\, dv'.
\label{6}
\end{equation}
Here, $P(v) = (dE/dt)/(dE/dt)_{\rm QF}$ is the rate at which the gravitational
waves remove orbital energy from the system, normalized to the quadrupole-formula
expression $(dE/dt)_{\rm QF} = -(32/5)(\mu/M)^2 v^{10}$; and $Q(v) = 
(dE/dv)/(dE/dv)_{\rm N} = (1-6v^2)/(1-3v^2)^{3/2}$ gives the differential relation
between orbital energy and orbital velocity, normalized to the Newtonian expression
$(dE/dv)_{\rm N} = -\mu v$. The functions $A(v)$ and $P(v)$ are calculated
by numerically integrating the Teukolsky equation for a small mass moving on a
circular orbit of the Schwarzschild black hole.

On the other hand, the templates are given by
\begin{equation}
\tilde{h}(f;\bbox{\lambda},\bbox{\theta}) \propto v^{-7/2}\, 
e^{i\psi(v;\bbox{\lambda},\bbox{\theta})},
\label{7}
\end{equation}
where $v^{-7/2}$ is the leading-order approximation to $A(v)$, 
and $\psi(v;\bbox{\lambda},\bbox{\theta})$ is a post-Newtonian approximation
to $\Psi(v)$. In Sec.~3, $\psi(v;\bbox{\lambda},\bbox{\theta})$ will be given by
the 2PN approximation obtained by Blanchet, Iyer, Will, and Wiseman \cite{BIWW} 
using post-Newtonian theory. In Sec.~4, $\psi(v;\bbox{\lambda},\bbox{\theta})$ will 
be given by the slow-motion, perturbation-theory results of Tanaka, Tagoshi, and 
Sasaki \cite{Tanaka}; this approximation does not incorporate the finite-mass-ratio 
terms of the true post-Newtonian series, but it is carried out to a high PN order.

It should be emphasized that the perturbation-theory results are valid 
only in the limit $\mu/M \to 0$. Nevertheless, I will still consider  
binary systems with large mass ratios, so as to give an indication of the
robustness of my conclusions. 

\section{Second post-Newtonian waveforms as search templates}

Serge Droz and I have calculated fitting factors for several binary
systems using 2PN waveforms as templates. Our results are presented in
the table, where for comparison I also display the fitting factors 
for Newtonian templates.

\begin{center}
\begin{tabular}{cccc}
\hline \hline
 & \multicolumn{2}{c}{advanced LIGO} & initial LIGO \\
System & FF--Newton & FF--2PN & FF--2PN \\
\hline 
& & & \\
$1.4 + 1.4\ M_\odot$ & $79.2 \% $ & $93.0 \% $ & $95.8\%$ \\
$0.5 + 5.0\ M_\odot$ & $51.6 \% $ & $95.2 \% $ & $89.7\%$ \\
$1.4 + 10\ M_\odot$ & $55.8 \% $ & $91.4 \% $ & $88.4\%$ \\
$10 + 10\ M_\odot$ & $70.1 \% $ & $91.7 \% $ & $86.3\%$ \\
$4 + 30\ M_\odot$ & $61.3 \% $ & $86.6 \% $ & $67.9\%$ \\
\hline \hline
\end{tabular}
\end{center}

We see that for the advanced version of LIGO and for most of the binary
systems considered, the fitting factors are all above the 90\% mark.
According to Apostolatos' criterion \cite{Apostolatos}, second 
post-Newtonian waveforms should make acceptably effective search 
templates. Additional details can be found in Ref.~\cite{DrozPoisson}.

\section{Pad\'e approximants}

Recently, Damour, Iyer, and Sathyaprakash (DIS) \cite{DIS} discovered a clever 
way of expressing the post-Newtonian waveforms in terms of Pad\'e approximants, 
which accelerates the convergence of the post-Newtonian series. Here I shall 
restrict all considerations to the test-mass limit, $\mu/M \to 0$. 

It is known that Pad\'e approximants work especially well for rational functions.
However, the orbital energy (per unit rest mass) of a test particle in Schwarzschild 
spacetime is not a rational function of the orbital velocity: $\tilde{E}(v) = 
(1-2v^2)/(1-3v^2)^{1/2}$. To circumvent this, DIS choose to employ the alternative 
energy function $\tilde{e}(v) \equiv  \tilde{E}^2 - 1 = -v^2(1-4v^2)/(1-3v^2)$, 
which is rational. In a post-Newtonian context, only the first few terms of a 
Taylor series about $v=0$ can be calculated. For example, at the 2PN level, 
$\tilde{e}(v) \simeq -v^2(1 - v^2 + 3 v^4)$. It is easy to check that the 
equivalent Pad\'e approximant coincides with the {\it exact} result. 

The Pad\'e series therefore converges rapidly to the exact result for
$\tilde{e}(v)$, and it may be used to estimate the position of the pole:
$v_{\rm pole} = 1/\sqrt{3}$. In the test-mass limit, this estimate is
exact. It is possible, however, that the estimate might not
be as accurate in the finite-mass-ratio case, for which an exact expression
is not available. As a way of challenging the DIS proposal, I will allow
for a possible uncertainty in the position of the pole, by introducing a 
free parameter $\xi$ such that $v_{\rm pole} = \xi/\sqrt{3}$.

The second element of the DIS proposal is to express $P(v)$, the luminosity 
function introduced in Sec.~2, also in terms of Pad\'e approximants. This 
requires some thought, because schematically, the post-Newtonian series 
for $P(v)$ (as calculated by Tanaka, Tagoshi, and Sasaki \cite{Tanaka}) 
takes the form
\begin{eqnarray}
P(v) &=& 1 + v^2 + v^3 + v^4 + v^5 + (1 + \ln v)v^6 
\nonumber \\ & & \mbox{}
+ v^7 + (1+\ln v)v^8 + (1+\ln v)v^9 
\nonumber \\ & & \mbox{}
+ (1+\ln v)v^{10} + (1+\ln v)v^{11};
\label{9}
\end{eqnarray}
the presence of logarithmic terms prevents a straightforward conversion
to a Pad\'e form. The way around this, as implemented by DIS, is to 
factorize these terms. DIS also choose to factorize a simple pole at 
$v=v_{\rm pole}$, because this pole, already present in the energy function, 
can be shown to appear also in the luminosity function. The new expression for 
$P(v)$ is therefore
\begin{eqnarray}
P(v) &=& \bigl[ 1 + (v^6 + \cdots + v^{11}) \ln v \bigr]
\nonumber \\ & & \mbox{} \times
\bigl[ 1 - v/v_{\rm pole} \bigr]^{-1}
\nonumber \\ & & \mbox{} \times
\bigl[ 1 + v + \cdots + v^{11} \bigr],
\label{10}
\end{eqnarray}
and DIS re-express the third factor in terms of its equivalent Pad\'e
approximant. 

Using the DIS proposal, I have calculated semi-maximized ambiguity functions 
(SMAF) for selected binary systems. The results depend on the order $n$ at which 
the post-Newtonian expansion is truncated. [For example, $n=6$ corresponds
to Eq.~(\ref{10}) with all terms of order $v^7$ and higher discarded;
this is the 3PN-Pad\'e approximation.] The results depend also on $\xi$, 
which parameterizes the possible uncertainty in the position of the pole. 
The values $\{1.0,0.9,1.1,\infty \}$ were selected; the choice $\xi = 
\infty$ corresponds to the absence of the second factor in 
Eq.~(\ref{10}) --- the pole is not factorized.

\begin{figure}[ht]
\special{hscale=30 vscale=30 hoffset=10.0 voffset=17.0
         angle=-90 psfile=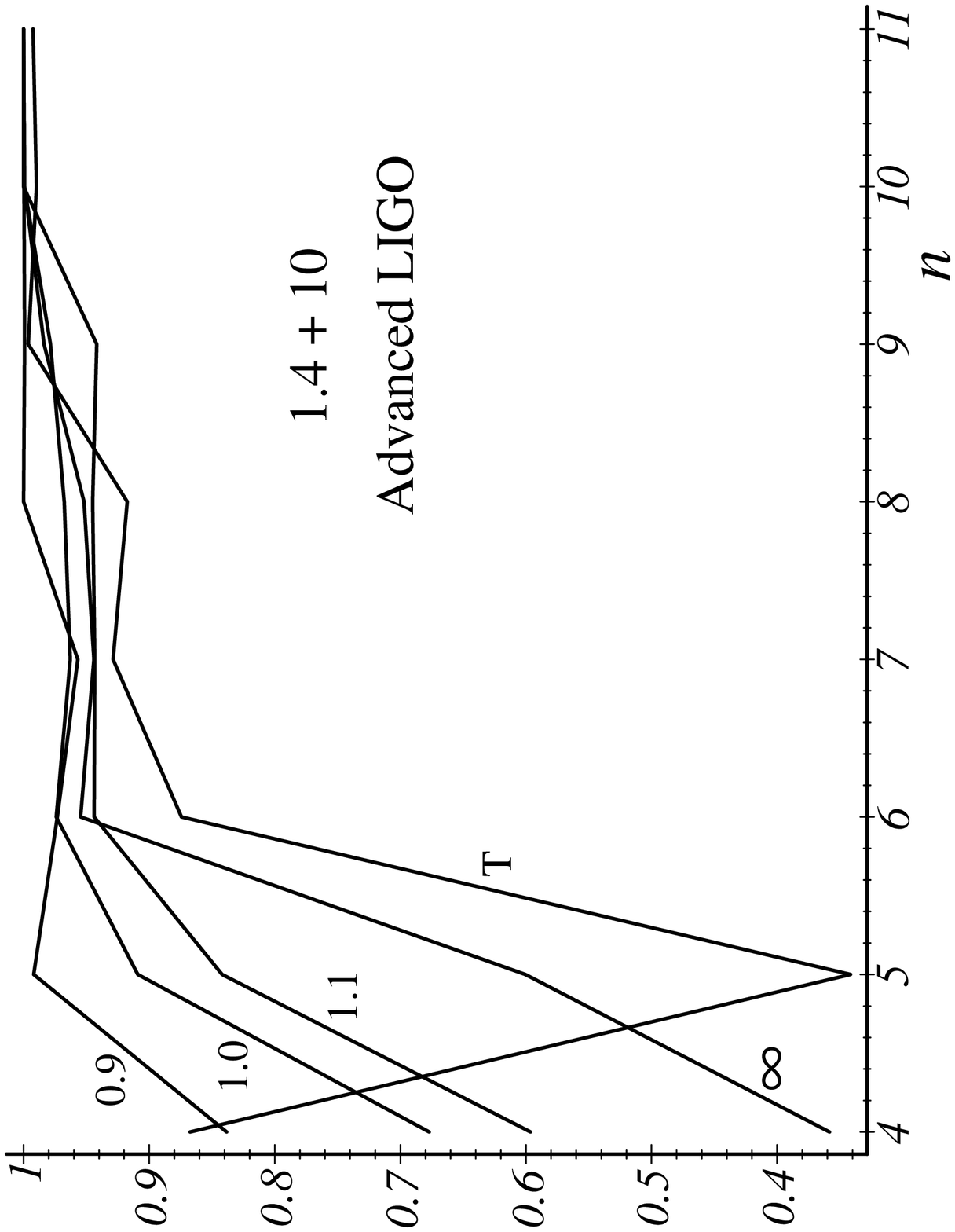}
\vspace*{2.0in}
\end{figure}

The representative case of a $1.4 + 10\ M_\odot$ binary system is 
represented in the figure, which plots the SMAF as a function of the
truncation order $n$, calculated using a noise curve appropriate
for an advanced LIGO detector. The curve labelled ``T'' is obtained
by using the usual Taylor expansion (\ref{9}) for the luminosity 
function, together with a Taylor expansion for $Q(v)$. The curves
labelled by numbers are obtained by using the Pad\'e form of 
Eq.~(\ref{10}), with $\xi$ given by the corresponding number. The
curve labelled ``$\infty$'' is also obtained by using the Pad\'e 
form of Eq.~(\ref{10}), but in the absence of the second
factor. The exact expression for $Q(v)$ is used for all the Pad\'e 
curves.

We see that the Pad\'e curves converge to unity much more rapidly 
than the Taylor curve. This demonstrates the great success of the DIS 
proposal. Furthermore, this is true for {\it all} the Pad\'e curves, 
whether or not they incorporate an uncertainty in $v_{\rm pole}$. I find 
this robustness of the DIS method quite remarkable. It is also interesting
to note that the choice $\xi = 0.9$ produces the {\it largest} values
for the SMAF. 

These results suggest that 3PN-Pad\'e waveforms will make a suitable
choice of search templates. They may also be sufficiently accurate for 
the reliable estimation of source parameters. Additional details will be 
presented elsewhere \cite{Poisson98}.

This work was supported in part by the Natural Sciences and
Engineering Research Council of Canada.

\end{document}